\documentclass[aps,pre,twocolumn,twoside,10pt]{revtex4-1}

\usepackage{graphicx,epstopdf,amsmath,amssymb,braket,color,epsfig,epstopdf,geometry}
\usepackage[colorlinks=true, citecolor=red, urlcolor=blue ]{hyperref}
\usepackage[table]{xcolor}
\usepackage{times}
\definecolor{red}{rgb}{0,0,0}

\begin{document}

\title{
Statistics of leading digits leads to unification of quantum correlations
} 
\author{Titas Chanda, Tamoghna Das, Debasis Sadhukhan, Amit Kumar Pal, Aditi Sen(De), and Ujjwal Sen}
\affiliation{Harish-Chandra Research Institute, Chhatnag Road, Jhunsi, Allahabad - 211019, India }

\begin{abstract}
 \textcolor{red}{We show that the frequency distribution of the first significant digits of the numbers in the data sets generated 
 from a large class of measures of quantum correlations, which are either entanglement measures, or belong to the 
 information-theoretic paradigm, exhibit a universal behavior. In particular, for Haar uniformly simulated 
 arbitrary two-qubit states, we find that the first-digit distribution corresponding to a collection of chosen computable 
 quantum correlation quantifiers tend to follow the first-digit law, known as the Benford's law, when the rank of the 
 states increases. Considering a two-qubit state which is obtained from a 
 system governed by paradigmatic spin Hamiltonians, namely, the XY model in a transverse field, and the XXZ model, 
 we show that entanglement as well as information theoretic measures violate the Benford's law. We 
 quantitatively discuss the violation of the Benford's law by using a violation parameter, and demonstrate that the violation parameter
 can signal quantum phase transitions occurring in these models. 
 We also comment on the universality of the statistics of first significant digits corresponding to 
 appropriate measures of quantum correlations in the case of multipartite systems as well as systems in higher dimensions.}
\end{abstract}

\maketitle

\section{Introduction}
\label{intro}

During the advancement of quantum information science, 
search for correlations having truly quantum nature has been in focus. 
Quantum entanglement \cite{ent_horodecki}, a manifestation of quantum correlation,  
has emerged 
as the key ingredient in several protocols related to quantum communication 
\cite{ASDUS,densecode,communication,teleport,diqc,crypto} and computation \cite{onewayQC}. 
Over time, exciting results such as non-zero non-classical efficiency of quantum states having vanishing entanglement, 
and locally indistinguishable orthogonal product states \cite{nl_w_ent,opt_det,dqci} have resulted in the development 
of quantum correlations beyond the ``entanglement-separability'' paradigm, 
broadly known as the quantum information-theoretic measures \cite{celeri_2011,modi_rmp_2012}. 
This has led to a large set of \textit{bona fide} measures of quantum correlations 
(see \cite{ent_horodecki,celeri_2011,modi_rmp_2012}, and the references therein), 
each of which, belonging usually to either of the two paradigms,
has its own degree of importance, applicability, and  computability.

The measures of quantum correlations, belonging to the entanglement-separability domain, are significantly different   
from those of the quantum information-theoretic origin, as indicated by their several properties
(see \cite{ent_horodecki,celeri_2011,modi_rmp_2012}, and the references therein). For example, 
entanglement measures are monotone under local operations and classical communication, 
whereas information-theoretic ones are not. Besides, a quantum state with vanishing entanglement can have a 
non-zero value of a chosen quantum information-theoretic measure, 
thereby putting the two of them on a different footing. 
Also, recent studies have shown that, under different types of environmental noise, entanglement  
exhibits a ``sudden death'' and vanishes at a finite time \cite{sd_group,sd_group2}, while   
a quantum information-theoretic measures decays asymptotically 
with time \cite{disc_dyn_group,disc_dyn_group2,disc_dyn_group3}, 
showing a more robust behavior against decoherence. It is also noteworthy that under specific initial conditions, 
quantum information theoretic measures may remain invariant over time \cite{disc_freez_group}, while 
the entanglement measures exhibit no such property (cf. \cite{ent_freez}). 

Interestingly, despite such differences, measures from the two domains show some similar behavior in a variety of 
physical scenarios. In the case of pure states, all the bipartite measures belonging to the two classes 
behave quite similarly \cite{ent_horodecki,modi_rmp_2012}.
Also, quantum correlation measures, irrespective of their origin, can be made monogamous by a
suitable choice of monotonically increasing function of the measure \cite{salini}. Moreover, 
measures from both the areas 
can be used to detect cooperative phenomena such as quantum phase transition 
(QPT) \cite{rev_xy,ent_xy_bp,ent_xy_mp,oth_mes_xxz,qpt_discord,qpt_book}, 
and the ``order-from-disorder'' phenomena \cite{order-from-disorder}. 
It is thus natural to ask whether an interlink exists between the values obtained from the  
different measures of quantum correlations having drastically different origins. 
There has been efforts to 
relate the measures of quantum correlations belonging to 
the two different paradigms \cite{unified,ent_disc,hsd_et_al}, 
although no conclusive result, as yet, exists. 

In this paper, we 
investigate whether the statistical properties of the different measures of quantum correlation, belonging to 
either of the paradigms, exhibit a universal behavior. 
Towards this aim, we analyze the frequency distribution of first-significant digits
of quantum correlations of both the classes obtained in different situations. An empirical law, 
known as the Benford's law,  
has been studied in varied fields including biology, geology, finance models, etc.
\cite{newcomb,benford,benfordweb,hill,busta,qpt_benford}, and states, in particular, that the first significant 
digit in any data is more often $1$. 
It has been shown that 
the law is  satisfied by several data sets obtained from various natural phenomena, while it can be violated in different 
physical systems including quantum spin models \cite{qpt_benford}.  
We show that the frequency distribution of the 
first significant digits 
occurring in a dataset corresponding to a measure of quantum 
correlation shows a decaying profile. 
We claim that a universal feature for all quantum correlation measures.
In particular, we demonstrate that the observed frequency distributions obtained from all the quantum correlations, 
in the case of arbitrary two-qubit states, tend to follow the Benford's law \cite{newcomb,benford,benfordweb},
and the violation parameter with respect to the Benford's law decreases with the increase of the rank of the 
quantum state.
Moreover, we observe that if deviation of Benford's law is quantified by distance measures like the  mean 
and standard deviations, and the Bhattacharya metric, the Benford violation parameter computed with respect to 
the Bhattacharya metric always possesses a lower value compared to the other distance metrics, irrespective of the 
quantum correlation measure and rank of the states.
We also consider the set of 
arbitrary two-qubit states in the parametric space, up to local unitary transformations, and verify the universal 
behavior of the first significant digits. The universality is found to be a low value of violation of the Benford's law, 
to be quantified below, is 
retained even when specific subsets of the 
set of two-qubit states are considered as the sample space for determining the first-digit distributions corresponding to
the quantum correlations. 

The universality may be hindered due to the constraints that are put to the quantum states
when the physical system is governed by a specific Hamiltonian. We investigate this by considering 
the well-known $XY$ model in a transverse external magnetic field \cite{xy_group}, 
defined on a system of $N$ spin-$\frac{1}{2}$ particles. We find 
that although the first-digit distribution does not mimic Benford's law, 
both in the case of entanglement and quantum information-theoretic measures except quantum work deficit, 
the Benford violation parameter can detect quantum phase transitions present in this model \cite{qpt_benford}. 
Interestingly, we observe that quantum discord changes its features of first-digit distribution qualitatively, by
changing specifically from having a decreasing to an increasing trend, when the system changes from paramagnetic 
to antiferromagnetic phases. \textcolor{red}{We also consider the $XXZ$ chain \cite{xxz_group,xxz_group_2}, and show that similar to the case of 
the $XY$ model, 
it is indeed possible to find a suitable quantum correlation measure whose Benford's violation parameter faithfully detects quantum 
critical points in this models.}

To observe the effect of increasing the number of parties and the dimensions of the systems, 
we also investigate the digit-distributions of distance-based entanglement and monogamy-based quantum correlation 
measures for three-qubit pure states, and computable quantum correlation measures for  
bipartite systems in higher dimensions. Specifically, we show that the observed frequency distribution of 
first significant digits obtained from the geometric measure of entanglement in the case of three-qubit systems can 
distinguish between two inequivalent classes of pure states, namely, the Greenberger-Horne-Zeilinger (GHZ) class
and the W class \cite{ghzstate,zhgstate,W-vs-GHZ,dvc}.

The paper is organized as follows. In Sec. \ref{blaw}, we provide a brief discussion on the Benford's law and 
the methodology adopted to compute the distribution of the first-significant digits. Sec. \ref{univ_corr} deals with
the discussions on the universal features of the first-digit distributions corresponding to different measures of 
quantum correlations. Sec. \ref{conclude} contains the concluding remarks.

\section{Statistics of leading digits: Newcomb-Benford's law}
\label{blaw}

The study of the statistics of leading digits started in 1881 \cite{newcomb}, when astronomer Simon Newcomb observed that
the occurrence of the digits $1$ to $9$ as the first significant digit of the numbers in a given set of data 
is not randomly distributed. Specifically, it was noticed that there exists many data sets for which the digit ``$1$'' occurs  
almost $30\%$ of the times, the digit ``$2$'' almost $17\%$, and the decreasing trend continues up to the digit ``$9$'', 
which appears almost $4.5\%$ of the times. The frequency distribution, $p_b(d)$, of the first significant digits, 
$d$ ($d\in\{1,2,\cdots,9\}$), is governed by an empirical law given by 
\begin{eqnarray}
 p_{b}(d)=\log_{10}\left(\frac{d+1}{d}\right).
 \label{ben_dist}
\end{eqnarray}
It is widely known as the Benford's law due to its rediscovery by Frank Benford in 1938, 
who verified the law for a wide range of natural datasets \cite{benford}.
Since then, frequency distribution of first significant digits of the numbers occurring in datasets of 
various origins has attracted a lot of attention of the scientific community, 
and the Benford's law has been tested in diverse areas of science \cite{benfordweb}.
Mathematical insight regarding the scale invariance of Benford's law has also been obtained in recent 
studies \cite{hill}.

\begin{figure}
 \includegraphics[scale=0.7]{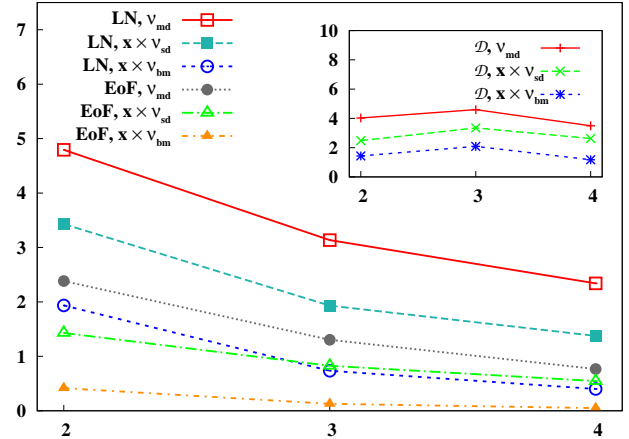}
 \caption{(Color online.) Variations of BVP (ordinate), namely,  
 $\nu_{md}$, $\nu_{sd}$, and $\nu_{bm}$, corresponding to LN and EoF with increasing rank, $r$ (abscissa),  
 in the case of arbitrary two-qubit states. (Inset) Trends of  
 $\nu_{md}$, $\nu_{sd}$, and $\nu_{bm}$ (ordinate), corresponding to $\mathcal{D}$, against $r$ (abscissa),  
 in the case of arbitrary two-qubit states. To compare the trends of the graphs, the values of $\nu_{sd}$ and $\nu_{bm}$
 are multiplied with a factor $x=50$ in the case of all the measures of quantum correlation. All quantities are 
 dimensionless.}
 \label{fig1}
\end{figure}

Interestingly, not all naturally occurring datasets obey Benford's law. The deviation of the observed 
frequency distribution, $p_o(d)$, can be quantified by computing the distance of $p_o(d)$ from $p_b(d)$. 
This quantity, known as the ``\textit{Benford violation parameter}'' (BVP), $\nu$ \cite{busta,qpt_benford}, 
depends on the two distributions as well as the distance-metric used to quantify the 
separation between them. For the present study, we consider three specific metrics, namely, 
the mean deviation (MD), the standard deviation (SD), and the Bhattacharya metric (BM), 
in terms of which the corresponding BVP are given by \cite{qpt_benford}
\begin{eqnarray}
\nu_{md}&=&\sum_{d=1}^{9}\left|\frac{p_{o}(d)-p_{b}(d)}{p_{b}(d)}\right|, \\
\nu_{sd}&=&\frac{1}{3}\left(\sum_{d=1}^{9}[p_{o}(d)-p_{b}(d)]^2\right)^{\frac{1}{2}}, \\
\nu_{bm}&=&-\ln\sum_{d=1}^{9}\left[p_{o}(d)p_{b}(d)\right]^{\frac{1}{2}},
\label{bvp}
\end{eqnarray}
where $\nu_{md}$, $\nu_{sd}$, and $\nu_{bm}$ correspond to MD, SD, and BM respectively. The concept of violation 
of Benford's law has been used in several scenarios as diverse as economics, election processes, digital image 
manipulation, seismology, and quantum phase transition \cite{busta,qpt_benford}.   

In the case of a data collected from a specific physical phenomenon, 
the measured quantity, $q$, usually has its own range of 
values, which might lead to a trivial violation of the Benford's law. For example, 
in a data corresponding to $q$ having values $2\leq q<3$, the first significant digit 
will always be $d=2$. 
Such triviality can be avoided by suitable shifting and scaling of $q$, achieved by 
the ``Benford's quantity'' (BQ), $q_{b}$, defined as 
\begin{eqnarray}
q_{b}=\frac{q-q_{min}}{q_{max}-q_{min}},
 \label{bquant}
\end{eqnarray}
where $q_{min}$ and $q_{max}$ are respectively the minimum and the maximum values of $q$ 
in the dataset. Eq. (\ref{bquant}) implies a mapping of the actual range of $q$ onto the range 
$[0,1]$ of $q_b$. Such a scaling also allows one to compare the different first-digit distributions obtained from 
quantities having different physical origins. Unless otherwise stated, we use BQ to compute the frequency distributions, 
$p_o(d)$, and the subsequent values of $\nu$.

\begin{table}
 \begin{tabular}{|c|c|c|c|c|c|c|}
 \hline
  BVP & $\mathcal{N}$ & LN & $\mathcal{C}$ & EoF & $\mathcal{D}$ & $W$ \\
 \hline 
  $\nu_{md}$ &  $1.770$  & $2.340$  & $2.431$ & $0.768$ & $3.497$ & $4.960$ \\
 \hline
  $\nu_{sd}\times10^2$ & $2.086$ & $2.751$ & $2.806$ & $1.091$ & $5.250$ & $6.231$ \\
 \hline 
  $\nu_{bm}\times10^3$ & $4.838$ & $8.026$ & $8.770$ & $0.982$ & $23.487$ & $43.074$ \\
 \hline  
 \end{tabular}
\caption{The values of $\nu_{md}$, $\nu_{sd}$, and $\nu_{bm}$ for $\mathcal{N}$, LN, $\mathcal{C}$, EoF, $\mathcal{D}$, 
and $W$ in the case of arbitrary two-qubit states of rank $4$.}
\label{tab}
\end{table}

\section{Universality in quantum correlations}
\label{univ_corr}
 
In the present study, we use two broad genres of measures of quantum correlations, namely, the entanglement measures, 
and the quantum information theoretic measures. Our choice of bipartite quantum correlations belonging to the 
first category include entanglement of formation (EoF) \cite{eform}, concurrence, $\mathcal{C}$ \cite{eform,mon}, 
negativity, $\mathcal{N}$, logarithmic negativity, 
$\mbox{LN}$ \cite{neg_group}, and relative entropy of entanglement, $S_R$ \cite{relent}, 
while in the second category, we take into account quantum discord (QD), $\mathcal{D}$ \cite{disc_group}, 
and one-way quantum work deficit (QWD), $W$ \cite{wdef_group}. 
In the multiparty cases, we restrict ourselves to pure states due to the lack of computable multipartite measures of 
quantum correlations in the case of mixed multiparty states, and also due to computational difficulties of numerical 
generation of multipartite mixed state. In this domain, we 
consider the generalized geometric measure (GGM),
$\mathcal{G}$ \cite{ggm}, and monogamy scores, $\delta$ \cite{salini,mon_ent,mon_disc,mon,score}, corresponding to 
squares of different bipartite quantum correlations, such as $\mathcal{C}$, $\mathcal{N}$, $\mathcal{D}$, and $W$.
Brief descriptions of these measures have been given in the Appendix.
We intend to investigate whether there exists any common feature to the 
frequency distributions of the digits occurring at the first significant position in the values of quantum correlations,
although they originated from widely different measures.

\begin{figure}
 \includegraphics[scale=0.7]{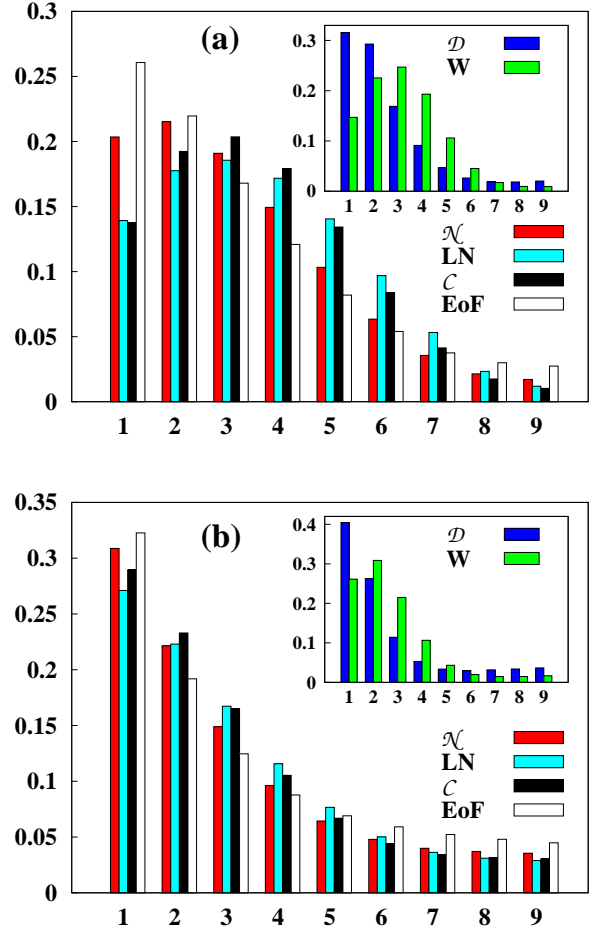}
 \caption{(Color online.) Histograms representation of  $p_o(d)$ (ordinate), 
 of the first significant digit, $d$ (abscissa), corresponding to $\mathcal{N}$, LN, 
 $\mathcal{C}$, and EoF, in the case of arbitrary two-qubit states of rank (a) $r=2$, and (b) $r=4$. The distributions 
 closely obey Benford's law for $r=4$, confirming the observation in Fig. \ref{fig1}. 
 The insets in the figures show the profiles of  
 $p_o(d)$ (ordinate) against $d$ (ordinate) in the case of quantum information-theoretic measures, viz., 
 $\mathcal{D}$ and $W$. All the quantities plotted are dimensionless.}
 \label{fig2}
\end{figure}

\subsection{Qubit systems}

We first consider the bipartite measures, and study the behavior of $p_o(d)$ in the case of quantum states in
$\mathbb{C}^2\otimes\mathbb{C}^2$. To compute $p_o(d)$,  
one can generate arbitrary pure as well as  mixed two-qubit 
states, Haar uniformly in the state space. Besides, one can also consider a parametrization of an arbitrary 
two-qubit state in terms of the correlation matrix and the local Bloch vectors. 
We consider both the cases in our analysis.

\subsubsection*{\textbf{Arbitrary two-qubit states}}
Let us first concentrate on the case of arbitrary two-qubit quantum states of different ranks.  
To determine $p_o(d)$ corresponding to different measures of quantum correlations in $\mathbb{C}^2\otimes\mathbb{C}^2$
systems, we Haar uniformly generate a sample of $10^6$ two-qubit states, $\rho_{12}$,  for  
each of the ranks $r=2$, $3$, and $4$. The rank-$1$ states form the set of pure states, 
on which we shall comment later. 
We find that the values of $\nu$, for different measures of quantum correlations, 
decreases with the increase of ranks of the states in the case of arbitrary two-qubit states. 
This is clearly observed from Fig. \ref{fig1}, where the variations of the values of $\nu$ (ordinates) with 
$r$ (abscissa) in the cases of 
$\mbox{LN}$, EoF, and $\mathcal{D}$ (in the inset) are shown 
for all the three distance metrics considered. We find that  
$\nu$ is minimum and considerably low for $r=4$, the maximum rank possible for a 
two-qubit state, implying a $p_o(d)$ very close to $p_b(d)$, irrespective of the distance measures.
Hence, it is reasonable to conclude that the frequency distribution of the first significant digits in a data 
corresponding to a given measure of quantum correlation closely mimics Benford's law when the two-qubit 
quantum states have full rank.  
Interestingly, we note that BVP for entanglement measures shows monotonic behavior with respect to the rank while 
that for quantum information-theoretic measures shows non-monotonicity with the increase of $r$. 
Fig. \ref{fig2}(a) and (b) depict the histogram representations of the profiles of $p_o(d)$ originated from 
different measures with $d$ in the cases of two-qubit mixed states of rank $r=2$ and $r=4$
respectively. 

Note that amongst the information-theoretic measures of quantum correlations, $p_o(d)$ corresponding to QD mimics
$p_b(d)$ more closely compared to that in the case of QWD, while $p_o(d)$ corresponding to the EoF have the least 
violation from the Benford's law among the entanglement measures considered in this study. 
The analysis shows that for any computable entanglement measure as well as information-theoretic measure, $\nu_{bm}$
possesses the minimum value among all the distance measures. For example, in the case of rank-$4$ states, 
we compare the values of $\nu$ obtained from different quantum correlation 
measures with different distance measures
(See Table \ref{tab}). For arbitrary two-qubit states, we therefore find a quantifier using Benford's law, 
where patterns unify all the 
quantum correlation measures, and erases their origin. On the other hand, violation parameter of Benford's law are 
capable to distinguish them.

\subsubsection*{\textbf{Two-qubit states in parameter space}} 

It is well-known that an arbitrary two-qubit state, 
up to local unitary transformation, can be expanded in terms of nine real parameters and the Pauli matrices,
$\sigma^{\alpha}$ ($\alpha=x,y,z$), as \cite{luo_pra_2008}
\begin{eqnarray}
 \rho_{12}&=&\frac{1}{4}\Big[I_{1}\otimes I_{2}
 +\sum_{\alpha=x,y,z}c^{\alpha\alpha}\sigma_{1}^{\alpha}\otimes\sigma_{2}^{\alpha}\nonumber \\
 &&+\sum_{\alpha=x,y,z}c_{1}^{\alpha}\sigma_{1}^{\alpha}\otimes I_{2}
 +\sum_{\beta=x,y,z}c_{2}^{\beta}I_{1}\otimes \sigma_{2}^{\beta}\Big].
 \label{twoqubitequiv}
\end{eqnarray}
Here, $c^{\alpha \alpha} = \langle\sigma^{\alpha} \otimes \sigma^{\alpha} \rangle$, called the ``classical 
correlators'', are the 
the diagonal elements of the correlation matrix with $|c^{\alpha\alpha}|\leq1$, 
$c_{1}^{\alpha} = \langle\sigma^{\alpha}_{1} \otimes I_{2}\rangle$ and 
$c_{2}^{\beta} = \langle I_{1} \otimes \sigma_{2}^{\beta} \rangle$, called the ``magnetizations'', 
are the elements of the two local Bloch vectors where $|c_1^{\alpha}|,|c_{2}^{\beta}|\leq1$, and
$I_{1(2)}$ is the identity operator in the Hilbert space of qubit $1(2)$. A two-qubit state $\rho_{12}$ of the 
form given in Eq. (\ref{twoqubitequiv}) can have a maximum rank, $r=4$. 

\begin{figure}
 \includegraphics[scale=0.7]{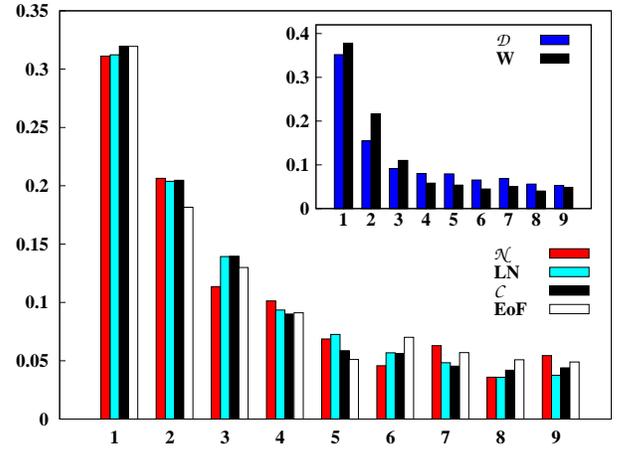}
 \caption{(Color online.) Histograms of $p_o(d)$ (ordinate) against $d$ (abscissa), 
 for the entanglement measures, $\mathcal{N}$, LN, 
 $\mathcal{C}$, and EoF, and the information-theoretic measures, $D$ and $W$ (inset) 
 for the two-qubit states of the form given in Eq. (\ref{twoqubitequiv}). 
 All the quantities plotted are dimensionless.}
 \label{fig3}
\end{figure}

One can generate random two-qubit states in the parameter space by choosing the relevant parameters in their 
allowed ranges from specific probability distributions to check the effect of different simulations of states on Benford's 
law. To compute $p_o(d)$ corresponding to different measures of quantum correlations, 
we simulate sets of $10^6$ random states of the form 
$\rho_{12}$ given in  Eq. (\ref{twoqubitequiv}) for each of the measures, 
by choosing the parameters from uniform distributions in appropriate ranges. We observe that $p_o(d)$, 
corresponding to all the quantum correlation measures, closely follow Benford's law, as indicated by 
the profiles of $p_o(d)$ depicted in Fig. \ref{fig3}.

Depending on the discussions presented above, it is natural to ask whether $p_o(d)$ corresponding to 
two-qubit states generated within specific subsets of the complete set of arbitrary two-qubit states  
possess such universal feature. To address this question, we consider three special instances, two of which
can be obtained as special cases of Eq. (\ref{twoqubitequiv}). 

\noindent\textbf{(i)} \textit{Bell-diagonal (BD) state.} It is obtained from 
Eq. (\ref{twoqubitequiv}) by setting $c_1^\alpha=c_2^\beta=0$, where $\alpha,\beta=x,y,z$. 
By generating a set of $10^6$ such random states for each of the 
quantum correlations, where the three diagonal 
correlators are drawn from a uniform distribution within their allowed ranges, we analyze the 
behavior of $p_o(d)$.

\noindent\textbf{(ii)} \textit{Two-qubit states with single magnetization.} Let us consider the 
state with $z$ magnetization
$(c_1^{x,y}=c_2^{x,y}=0,c_1^z\neq c_2^z\neq0)$. Here, $\rho_{12}$, written in the computational basis 
$\{|00\rangle,|01\rangle,|10\rangle,|11\rangle\}$, assumes the form of an $X$-state \cite{xstate}, given by
\begin{equation}
\rho_{12}=
\left(
\begin{array}{cccc}
   a_{1}&0&0&b_{1}\\
   0&a_{2}&b_{2}&0\\
   0&\bar{b}_{2}&a_{3}&0\\
   \bar{b}_{1}&0&0&a_{4}
\end{array}
\right),
\label{X}
\end{equation}
where the matrix elements are real ($b_1=\bar{b}_1$, $b_2=\bar{b}_2$). 
Similar to the previous case, here also we study the trends of 
$p_o(d)$ over a set of $10^6$ random states for each of the measures of quantum correlations. 
However, in the present case, the sets are generated by 
choosing the matrix elements from uniform and normal distributions, 
in such a way that $\rho_{12}$ becomes a valid density matrix.

\noindent\textbf{(iii)} \textit{Generic two-qubit $X$ state.} The state is of the form given in Eq. (\ref{X}), 
but with complex off-diagonal elements. In this case, a set of $10^6$ random states is generated by choosing the real 
and imaginary parts 
of the matrix elements from uniform and normal distributions.

In all the three cases listed above, we observe that $p_o(d)$, for both entanglement and information-theoretic measures, 
closely follow Benford' law. 
For example, we find that the BVP, $\nu_{bm}$, corresponding to EoF, LN, $D$, and $W$ are respectively
$3\times10^{-4}$, $3.2\times10^{-3}$, $2.7\times10^{-3}$, and $4.5\times10^{-3}$ for the $X$ states,
when the matrix-elements are real, and are chosen from normal distributions. 
In the case of the BD states, we compute $p_o(d)$ corresponding to the relative entropy 
of entanglement, which, in addition to all the other quantum correlation measures, obeys such universality.   
Hence the universality of the variation of $p_o(d)$ against $d$ is retained in 
the case of entanglement measures as well as quantum information-theoretic measures,
when arbitrary two-qubit states in the parameter space are considered.
Such analysis indicates that the quantum correlation measures, irrespective of the choice of the measures, tend to follow 
Benford's law, especially when the states are of full rank, independent of the simulation process. 

\begin{figure}
 \includegraphics[scale=0.705]{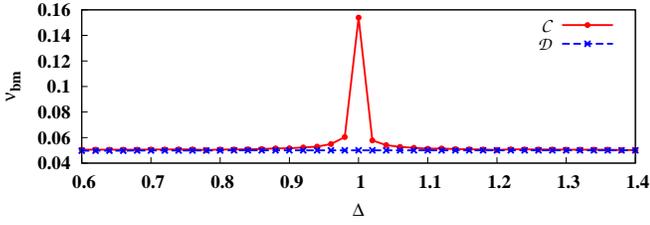}
 \caption{(Color online.) Variations of $\nu_{bm}$ corresponding to $\mathcal{C}$ and $\mathcal{D}$
 as functions of $\Delta$ in the 
 case of XXZ model with $N=8$. The ground state of the model is determined by exact diagonalization technique. 
 All quantities plotted are dimensionless.}
 \label{fig4}
\end{figure}

\begin{figure}
 \includegraphics[scale=0.7]{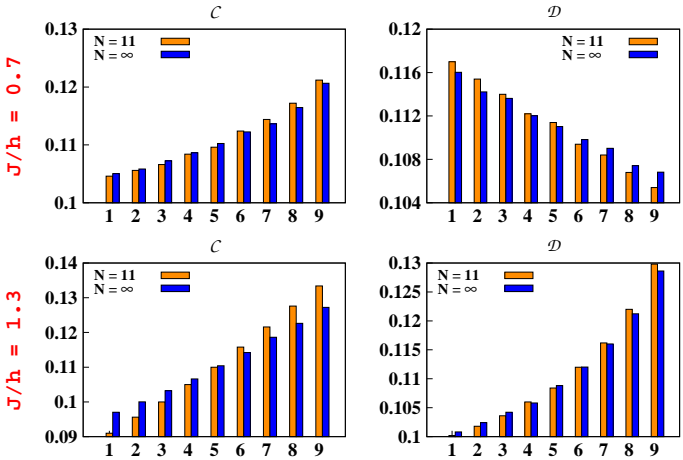}
 \caption{(Color online.) Variations of $p_o(d)$ (ordinate) against $d$ (abscissa) in the case of $\mathcal{C}$ (left panels) 
 and $\mathcal{D}$ (right panels). The two-qubit states are the zero-temperature states of the transverse-field 
 $XY$ model with $J/h=0.7$ (top panels) and $J/h=1.3$ (bottom panels). All quantities are dimensionless.}
 \label{fig5}
\end{figure}

\begin{table}
\begin{tabular}{|c|}
 \hline
 $\nu_{bm}\times 10^2$ \\
  \begin{tabular}{c|c|c|c|c|c|c}
 \hline
  $J/h$ & $\mathcal{N}$ & LN & $\mathcal{C}$ & EoF & $\mathcal{D}$ & $W$ \\
 \hline 
   $0.7$ & $5.587$ & $5.656$ & $5.606$ & $5.305$ & $4.593$ & $3.875$   \\
 \hline  
   $1.3$ & $5.627$ & $5.700$ & $6.378$ & $4.937$ & $6.102$ & $4.142$   \\
 \end{tabular}\\
 \hline  
\end{tabular}
\caption{The values of $\nu_{bm}$ for $\mathcal{N}$, LN, $\mathcal{C}$, EoF, $\mathcal{D}$, 
and $W$ in the case the zero-temperature states of the transverse-field XY model with $J/h=0.7$ and 
$J/h=1.3$ in the thermodynamic limit.}
\label{tab2}
\end{table}

\subsubsection*{\textbf{Physical Systems: Quantum Spin Models}}

\textcolor{red}{In all the qubit systems considered so far in this paper, we have generated quantum states uniformly in either 
the state space, or the parameter space of the corresponding states. 
In all these cases, we have seen that quantum correlation measures almost follow the Benford's law and can not detect the 
generation process of the states. 
However, there exist physical systems in which the allowed quantum states 
are governed by the Hamiltonian describing the system. It is, thus, natural to ask what 
happens to such unifying features of 
$p_o(d)$, corresponding to different quantum correlations, if it is
constrained by the system Hamiltonian.
To address this question, we consider a  quantum spin model in one dimension, 
given by the Hamiltonian
\begin{eqnarray}
H&=&\frac{J}{2}\sum_{i=1}^N \left\{(1+\gamma)\sigma_i^x\sigma_{i+1}^{x}+(1-\gamma)\sigma_i^y \sigma_{i+1}^{y}
+\Delta\sigma_i^z\sigma_{i+1}^z\right\}\nonumber\\&&
+h\sum_{i=1}^N \sigma_i^z.
\end{eqnarray} 
Here, $J$, and $h$ are respectively the strengths of the exchange interaction and the transverse magnetic field. 
The anisotropies in the strengths of the exchange interaction are given by  $\Delta$ in the $z$ direction, and $\gamma$ in the $x − y$ 
direction. The total number of quantum spin-$1/2$ particles in the system is $N$.
We consider two specific cases of the Hamiltonian $H$: \textbf{(a)} the anisotropic XY model in a transverse field, 
represented by $\Delta=0$ \cite{xy_group}, and \textbf{(b)} the $XXZ$ model, given by $\gamma=0$ and $h=0$ \cite{xxz_group}. 
In the case \textbf{(a)}, with a choice of the system parameter as $\lambda\equiv J/h$, the model, with increasing $\lambda$, undergoes a 
QPT at $\lambda=1$ from an antiferromagnetic phase to a quantum paramagnetic phase \cite{qpt_book}.
In the case of \textbf{(b)} with $\lambda\equiv\Delta$, 
two QPTs take place at $\Delta=\pm1$. 
For $\Delta=-1$, the ground state of the model undergoes a Kosterlitz-Thouless (KT) QPT ,
while at $\Delta=1$, a QPT from the metalic phase $(0\leq\Delta\leq1)$ to the insulating phase $(\Delta>1)$ occurs \cite{xxz_group,xxz_group_2}.
}

\textcolor{red}{For the purpose of the present study, we are interested to look at the profile of $p_o(d)$ of the zero-temperature state  
in the different phases of the model. 
Note here that the present case is an example where the observable, $q$, may have an allowed 
range of values in the corresponding phase, and thus may result in a trivial violation of the Benford's law, 
as discussed in Sec. \ref{blaw}. To avoid such trivial violations, we employ the following procedure \cite{qpt_benford}. 
Let us assume that in a particular phase, $q$ is bounded in the range $[q_1,q_2]$. We consider a fixed value of the 
system parameter, $\lambda=\lambda_{0}$, in the given phase, and choose a small interval around $\lambda_0$, of width 
$\epsilon$, given by $(\lambda_0-\epsilon/2,\lambda_0+\epsilon/2)$. 
In this interval, we sample $n$ values of $\lambda$. Corresponding to those $n$ values of $J/h$,  
we create a data of $n$ values of $q$, in which the range of 
$q$ is given by $[q_1^\prime,q_2^\prime]$. Next, we scale the observable, $q$, so that all of the $n$ data points in 
$[q_1^\prime,q_2^\prime]$ now lie in the range $[0,1]$. This is achieved by a scaling similar to 
that described in Eq. (\ref{bquant}), where $q_{min}=q_1^\prime$, and $q_{max}=q_2^\prime$, and the scaled data now 
belongs to the corresponding BQ, $q_b$. The next step is to determine the frequency 
distribution of first significant digits from the data of $q_b$, 
which provides the $p_o(d)$ corresponding to the system parameter $\lambda=\lambda_0$. 
}

\textcolor{red}{The QPTs in both the models have been detected by standard condensed matter physics techniques \cite{qpt_book,xy_group,xxz_group} 
as well as several quantum correlation measures (see \cite{rev_xy}, and references herein). It has been shown that BVPs corresponding 
to transverse magnetization and two-site LN can also signal the occurrance of the QPT in the transverse-field XY model \cite{qpt_benford}. 
In the present study, we find that for the XXZ model, all the QPTs on the $\Delta$ axis described above are signalled by the BVP 
corresponding to different entanglement measures even when the system size is small, but not by the same corresponding to information theoretic measures. 
Fig. \ref{fig4} depicts the variation of $\nu_{bm}$ correponding to nearest neighbour concurrence and quantum discord for the ground state
of the $XXZ$ model with $\Delta$. The QPT at $\Delta=1$ is signaled by a sharp kink in the case of concurrence, while no change in the 
case of quantum discord takes place.}

\textcolor{red}{In the case of the $XY$ model, we choose two specific values of the system parameter $\lambda$, given by $\lambda=0.7$ and $\lambda=1.3$, 
in two separate phases of the transverse-field $XY$ model, and determine $p_o(d)$ corresponding to different 
bipartite measures of quantum correlations for the zero-temperature case of the model. 
In each case, $p_o(d)$ is computed from a sample of $n=5\times10^3$ data 
points in the vicinity of the fixed value of $\lambda$, with a small fixed interval $\epsilon=0.1$. 
Curiously, we observe that although the profile of $p_o(d)$, in the case of entanglement measures, 
show a complementary behavior to that of $p_b(d)$, the universality in the behavior of $p_o(d)$, 
in the sense that the BVP being small throughout, 
remains invariant 
across the QPT. On the other hand, the behavior of $p_o(d)$ corresponding to QD changes across the QPT, similar to 
the transverse magnetization \cite{qpt_benford}, while 
that corresponding to QWD remains unchanged. 
Although BVP for entanglement measures can detect QPT, the analysis of $p_o(d)$ reveals that the distribution trend of 
digits for QD clearly differentiates the two phases, which is not possible by considering entanglement measures.
Moreover, the Benford satisfying nature of the digit distribution, 
$p_o(d)$, in the sense that both $p_o(d)$ and $p_b(d)$ have decaying profiles with respect to the first significant
digits, for any quantum correlation measure, showing universality among quantum correlation measures, observed for 
arbitrary two-qubit states as well as restricted classes of states, is washed out when the two-qubit state is generated 
from a given Hamiltonian. 
The profiles of $p_o(d)$ against $d$, for $\mathcal{C}$ as the 
entanglement measure and QD as the quantum information theoretic measure in the different 
phases of the transverse-field $XY$ model, are shown in Fig. \ref{fig5}. One must note here that even a finite sized
system, with size as small as $N=11$, is sufficient to mimic the system in the 
thermodynamic limit ($N\rightarrow\infty$). An important point to note is that among all the entanglement as well as 
information theoretic measures, the frequency distribution corresponding to QWD, 
originated from the ground state of the model, satisfies Benford's law most closely. 
This is clearly visible from the values of 
$\nu_{bm}$ computed from $p_o(d)$ obtained from different measures of quantum correlations in the two different 
phases of the model (see Table \ref{tab2}).}

\subsection{Universality in higher dimensions and higher number of parties}
\label{multipart}


It is now natural to ask 
whether the universal behavior of $p_o(d)$ is generic when systems with higher number of parties, or in higher 
dimensions are considered. However, generating multipartite quantum states as well as states in higher dimensions 
having all possible ranks is itself a non-trivial problem. For the sake of completeness, in our study, we 
comment on the nature of $p_o(d)$ in the cases of certain paradigmatic classes of states. More specifically, we consider 
pure states in three-qubit systems, the complete set of which is 
constructed by the union of two independent classes of states, namely, the GHZ class, and the W class 
\cite{ghzstate,zhgstate,W-vs-GHZ,dvc}.  
In the case of higher-dimensional systems, we simulate
rank-$2$ states in $\mathbb{C}^2\otimes\mathbb{C}^3$  and $\mathbb{C}^2\otimes\mathbb{C}^4$. 

To check for 
unifying features as 
in the case of two-qubit systems, we Haar uniformly generate sets of $10^6$ states in each of the cases 
including tripartite pure states of GHZ and the W classes, and higher dimensional states in 
$\mathbb{C}^2\otimes\mathbb{C}^3$ and $\mathbb{C}^2\otimes\mathbb{C}^4$ systems, 
and determine the relevant quantum correlations. See Appendix for brief descriptions of the GHZ class, the W class,
and different quantum correlation measures considered.
It is important to note that the measures that we consider for two-qubit systems do not 
have a direct generalization for three-qubit systems, and hence we choose a distance-based entanglement measure, namely, 
GGM \cite{ggm}, and monogamy based measures \cite{mon,mon_disc,mon_ent} originated from bipartite quantum correlations. 

\begin{figure}
 \includegraphics[scale=0.7]{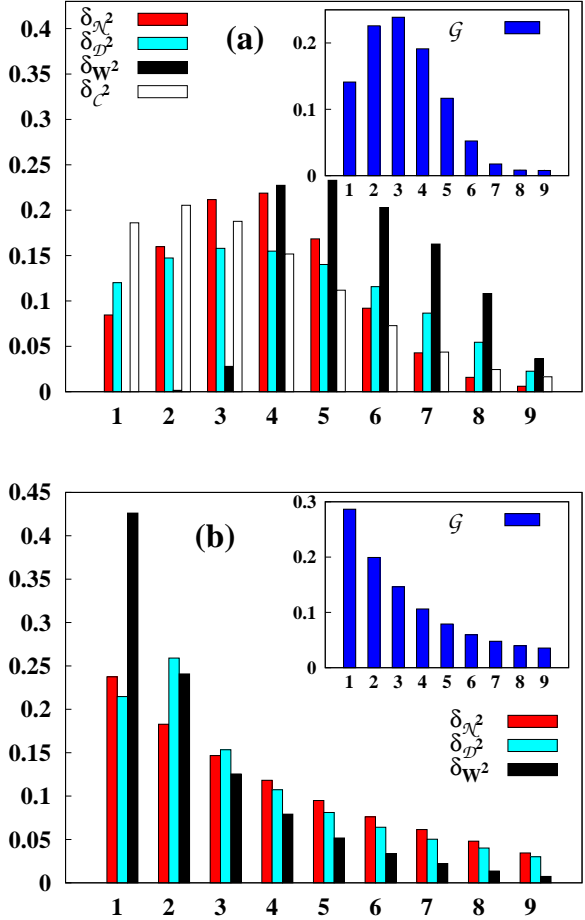}
 \caption{(Color online.) Profiles of $p_o(d)$ (ordinate) with $d$ (abscissa) in the case of multipartite measures 
 of quantum correlations. The measures that we choose are 
 $\delta_{\mathcal{N}^2}$, $\delta_{\mathcal{C}^2}$, $\delta_{\mathcal{D}^2}$, 
 and $\delta_{W^2}$ for arbitrary three-qubit states belonging to (a) GHZ and (b) W classes. The insets 
 show the corresponding variations of $p_o(d)$ (ordinate) with $d$ (abscissa) when GGM is chosen to be the 
 measure of quantum correlation. All quantities are dimensionless.}
 \label{fig6}
\end{figure}

The profiles of $p_o(d)$ and the corresponding values of
$\nu$ for different quantum correlations, in all the mentioned cases, are found to be inconclusive in 
determining whether such universality exists. Note that the multiparty states considered here 
are pure (rank-$1$) states, while all the higher-dimensional bipartite states have rank-$2$,  and
in the case of qubit systems, the universal feature becomes prominent in 
the case of states with higher rank. 
Hence it is reasonable to infer that conclusive result on the universality 
of the behavior of $p_o(d)$ in the case of multipartite systems as well as systems in higher dimensions 
can only be obtained once the corresponding states up to their full rank can be considered. 
The behavior of $p_o(d)$ against $d$ for various measures of quantum correlations in the case of three-qubit 
pure states are shown in Fig. \ref{fig6} for the GHZ (Fig. \ref{fig6}(a)) and the W (Fig. \ref{fig6}(b)) classes. 
For the three-qubit pure states, $p_o(d)$ corresponding to the W class 
exhibit considerable unifying behavior.  For example, the violation $\nu_{bm}$, obtained from the simulation of 
the W-class states 
for $\mathcal{G}$, $\delta_{\mathcal{N}^2}$, 
$\delta_{\mathcal{D}^2}$, and $\delta_{W^2}$ are respectively given by $2.006\times10^{-3}$, 
$3.832\times10^{-3}$, $9.846\times10^{-3}$, and $31.593\times10^{-3}$, showing 
a very low Benford's violation similar to the two-qubit states. Here, the monogamy scores corresponding to $\mathcal{D}^2$
and $W^2$ are computed by performing measurement over the nodal observer. Interestingly, as seen in Fig. (\ref{fig6})(a), 
the distribution, $p_o(d)$, for all quantum correlation measures 
obtained from the GHZ class do not closely follow Benford's law.
Therefore, while the value of GGM can not 
distinguish these two inequivalent classes, its first-digit distribution can. 

\section{Concluding remarks}
\label{conclude}

In this paper, we showed that statistics of the first significant digits occurring in the data 
corresponding to different measures of quantum correlations, belonging to both entanglement-separability as well 
as the information-theoretic paradigms, exhibit a universal feature. More specifically, according to our data-analysis,  
the frequency distribution of the first significant digits in the numbers corresponding to 
any \textit{bona fide} measure of quantum correlation tend to obey Benford's law, 
when states with higher ranks are considered, irrespective of their origin and nature, in general.
This feature becomes prominent when 
quantum states with full rank are considered. 
\textcolor{red}{We also discuss the effect when states corresponding to specific Hamiltonians, namely, the $XY$ 
model in a transverse field, and the XXZ model are considered.   
We find that although the universality is washed out 
in such cases,  BVP  can detect QPTs occurring in these models.
Such investigation revealed that in the case of the $XY$ model, 
the first-digit distribution of QD in the paramagnetic phase follow the Benford's 
law more closely than in the antiferromagnetic phase, and therefore digit distributions of QD
clearly identifies the quantum phases. 
On the contrary, concurrence and logarithmic negativity almost equally  
violate the Benford's law in both the phases, although 
BVP can still detect the QPT present in the model.} 
We finally analyze first significant digit distribution of different computable 
quantum correlation measures for multipartite as well as higher-dimensional systems. 
Specifically, in a tripartite domain, 
the first-digit distribution of geometric measure of entanglement for the W-class states closely follow Benford's law, while it 
is not the case for the GHZ class, indicating another quantity of completely different root that can identify these 
two classes. 

\appendix

\section*{Appendix}

\section*{Measures of Quantum Correlation}
\label{measures}

In this section, we briefly discuss the quantum correlation measures used in this study. We mainly consider two 
distinct classes of measures, namely, entanglement measures \cite{ent_horodecki}, and information-theoretic
measures of quantum correlations \cite{celeri_2011,modi_rmp_2012}.

\subsection*{Bipartite systems}

There is a variety of bipartite measures of quantum correlations existing in literature, 
each having their own degree of utility and computability. Here, we briefly discuss only a few selected measures 
relevant for the present study.

\subsubsection*{\textbf{Entanglement measures}}

The set of entanglement measures (see \cite{ent_horodecki}, and the references therein) considered here include 
von Neumann entropy as a measure of pure state entanglement \cite{srho}, 
and entanglement of formation \cite{eform}, concurrence \cite{eform,mon}, negativity, and logarithmic 
negativity \cite{neg_group} as the entanglement measures for mixed bipartite quantum states.

\noindent\textit{von Neumann entropy:} The von Neumann entropy, $S({\rho})$, of a quantum 
state $\rho$, is defined as $S(\rho) = - \mbox{tr}[\rho \log_2 \rho]$.  In the case of a pure 
bipartite quantum state $|\psi\rangle_{ab}$, shared between two parties ``$a$'' and ``$b$'', entanglement
between $a$ and $b$ is quantified as $S(\rho_{a/b})$ \cite{srho}, where $\rho_{a(b)}=\mbox{tr}_{b(a)}[\rho_{ab}]$ 
is obtained from $\rho_{ab}=|\psi\rangle_{ab}\langle\psi|$ by tracing out the party $b(a)$. 

\noindent\textit{Entanglement of formation and concurrence:} 
The definition of concurrence originates from that of entanglement of formation (EoF) \cite{eform} of a 
bipartite system, which is defined 
as the amount of singlets, $|\psi^-\rangle=(|01\rangle+|10\rangle)/\sqrt{2}$, 
required to prepare the quantum state of the system
by  local operations and classical communications (LOCC). 
For a bipartite mixed state $\rho_{ab}$, EoF is provided by its convex roof definition 
$E(\rho_{ab})=\min\sum_{i}p^iE(|\psi^i\rangle_{ab})$, where the minimization is taken over all possible pure 
state decompositions of $\rho_{ab}$, such that $\rho_{ab}=\sum_{i}p^iP[|\psi^i\rangle_{ab}]$. 

The optimization involved in the definition of EoF for a bipartite mixed state is what makes the measure 
intractable. However, for an arbitrary bipartite two-qubit state $\rho$, EoF can be obtained as a monotonic 
function of concurrence, $\mathcal{C}$, given by  \cite{eform,mon}
\begin{eqnarray}
\mathcal{C}=\text{max}\{0, \lambda_1-\lambda_2- \lambda_3-\lambda_4\},
\end{eqnarray} 
where $\lambda_i^2$'s, $i=1,\cdots,4$, are eigenvalues of the matrix $\rho\tilde{\rho}$ in decreasing order, with 
$\tilde{\rho}=(\sigma_y \otimes \sigma_y)\rho^*(\sigma_y \otimes \sigma_y)$. Here, $\sigma_y$ is the Pauli matrix, and 
the complex conjugation in $\rho^*$ is carried out in the computational basis. 

\noindent\textit{Negativity and logarithmic negativity:} 
Based on the Peres-Horodecki separability criterion \cite{neg_part_group}, negativity \cite{neg_group}, $\mathcal{N}$, for a 
bipartite quantum state, $\rho_{ab}$, is defined as 
\begin{eqnarray}
\mathcal{N}(\rho_{ab}) = \frac{||\rho_{ab}^{T_{a}}||-1}{2},
\end{eqnarray} 
where $||\varrho|| = \mbox{Tr}\sqrt{\varrho^{\dagger}\varrho}$ is the trace norm of the matrix $\varrho$, and 
$\rho_{ab}^{T_{a}}$ is obtained by performing partial transposition of $\rho_{ab}$ w.r.t. the party $a$ \cite{neg_part_group}. 
The logarithmic  negativity \cite{neg_group}, LN, can be obtained from $\mathcal{N}$ as
\begin{eqnarray}
\mbox{LN} = \log_2\left(2 \mathcal{N} + 1 \right).
\end{eqnarray}

\noindent\textit{Relative entropy of entanglement.} The relative entropy of entanglement, $S_R$, is a measure 
of entanglement quantifying the minimum relative entropy distance of an entangled state, $\rho$, from 
the set of separable states, $\mathcal{S}=\{\sigma\}$. It is given by 
\begin{eqnarray}
 \mathcal{S}=\underset{\sigma\in\mathcal{S}}{\min}S(\rho||\sigma),
\end{eqnarray}
where $S(\rho||\sigma)$ is the quantum relative entropy, given by 
$S(\rho||\sigma)=\mbox{tr}\Big(\rho\log_2\rho-\rho\log_2\sigma\big)$.

\subsubsection*{\textbf{Quantum information theoretic measures}} 

The set of quantum information-theoretic measures of quantum correlations considered in this paper include 
quantum discord and one-way quantum work deficit (see \cite{celeri_2011,modi_rmp_2012}, and the references therein). 

\noindent\textit{Quantum discord:} The total correlation between the 
two parties of a bipartite system is quantified by the quantum mutual 
information \cite{total_corr}. 
The definition of quantum discord of a bipartite quantum system $\rho_{ab}$ emerges from the 
difference between the quantum extensions of two equivalent ways to define mutual information in the classical domain. 
In its original form, quantum discord is defined as \cite{disc_group} 
\begin{eqnarray}
 \mathcal{D}_{a}=S(\rho_{a})-S(\rho_{ab})+\underset{\{\Pi_a\}}{\min}\sum_{i}S(\rho_{ab}^i),
\end{eqnarray}
where $S(\rho_{ab}^i)$ is the von Neumann entropy of the state 
$\rho_{ab}^i=M^i_{a}\rho_{ab}M^i_{a}/p^i$, obtained with probability $p^i$, from the projection-valued measurement, 
$\Pi_{a}^i$, performed over the subsystem $a$. Here, $M_{a}=\Pi_{a}^i\otimes\mathbb{I}_{b}$, 
$p^i=\mbox{Tr}[M^i_{a}\rho_{ab}M^i_{a}]$, and $\mathbb{I}_{b}$ is the identity operator in the Hilbert space of $b$. 
The subscript ``$a$'' in $\mathcal{D}_{a}$ denotes that the measurement is performed over the subsystem ``$a$''. 
Note here that the definition of quantum discord has an inherent asymmetry due to the local projective measurement 
performed over the subsystem ``$a$''. In general, $\mathcal{D}_a\neq \mathcal{D}_b$.

\noindent\textit{Quantum work deficit:} 
The one-way quantum work deficit (QWD) is defined as the difference between the amount of pure states extractable 
under global ``closed operation" (CO),  and ``closed local operations and classical communications" (ClOCC)
\cite{wdef_group}. 
The set of CO  consists of \textit{(i)} global unitary operations, and \textit{(ii)} dephasing by the 
set of projectors defined on the Hilbert space of $\rho_{ab}$, while the class of CLOCC consists of 
{\it (i)} local unitary operations, {\it (ii)} dephasing by local measurement on the subsystem $a$ ($b$), and 
{\it (iii)} communicating the dephased subsystem to the other party, $b$ ($a$), over a noiseless quantum channel.
It can be shown that the amount of pure states extractable from $\rho_{ab}$ under CO is given by 
\begin{eqnarray}
I_{\mbox{\scriptsize CO\normalsize}}(\rho_{AB})=\log_{2}\mbox{dim}\left(\mathcal{H}\right)-S(\rho_{AB}), 
\label{co}
\end{eqnarray}
while the same under CLOCC is given by 
\begin{eqnarray}
I_{\mbox{\scriptsize CLOCC\normalsize}}=\log_{2}\mbox{dim}\left(\mathcal{H}\right)-
\underset{\left\{\Pi_{a}\right\}}{\min}
S\left(\rho_{ab}^{\prime}\right).
\end{eqnarray}
where $\rho_{ab}^{\prime}=\sum_{i} p^{i}\rho^i_{ab}$ is the average quantum 
state after the projective measurement $\{\Pi_{a}\}$ on $a$. The QWD, $W$, is then defined as 
$W = I_{\mbox{\scriptsize CO\normalsize}}\left(\rho_{ab}\right) 
- I_{\mbox{\scriptsize CLOCC\normalsize}}\left(\rho_{ab}\right)$.

\subsection*{Multipartite systems}

The set of multipartite measures considered here include the generalized geometric measure (GGM)
\cite{ggm}, and monogamy scores \cite{salini,mon_ent,mon_disc,score} 
corresponding to different measures of quantum correlations. 

\noindent\textit{Generalized geometric measure:} 
Based on the geometric measure of entanglement \cite{geom_ent,geom_ent_ksep,geom_ent_book}, 
the generalized geometric measure (GGM) \cite{ggm} 
is a quantifier of 
genuine multipartite entanglement content in a multiparty pure state. 
A pure quantum state, $|\psi\rangle_{a_{1}a_{2}\cdots a_{n}}$, shared between $n$ parties given by 
$a_1,a_2,\cdots,a_n$, is said to be genuinely multiparty entangled if it can not be written as a product across any 
bipartition. The GGM of a pure quantum state, $|\psi\rangle_{a_{1}a_{2}\cdots a_{n}}$, is defined as
\begin{eqnarray}
\mathcal{G}(|\psi\rangle_{a_{1}a_{2}\cdots a_{n}})=1-\max|\langle\phi|\psi\rangle|^2_{a_{1}a_{2}\cdots a_{n}},
\end{eqnarray}
where the maximization is taken over all $n$-partite pure states $\{|\phi\rangle_{a_{1}a_{2}\cdots a_{n}}\}$, 
that are not genuinely 
$n$-party entangled. The advantage in using GGM as a quantifier of pure-state entanglement is that for 
multipartite pure states in arbitrary dimensions, $\mathcal{G}$ can be calculated as 
\begin{eqnarray}
\mathcal{G}(|\psi\rangle_{a_{1}a_{2}\cdots a_{n}}) = 1-\max\{\lambda^2_{{\cal A}: {\cal B} }\},
\label{ggmdef}
\end{eqnarray}
where $\lambda_{{\cal A}: {\cal B}}$ is the maximal Schmidt coefficient of  
$|\psi\rangle_{a_{1}a_{2}\cdots a_{n}}$ in the ${\cal A}: {\cal B}$ bipartite split, 
provided ${\cal A} \cup {\cal B} = \{a_{1},a_{2},\cdots,a_{n}\}$, and  ${\cal A} \cap  {\cal B} = \emptyset$. 
The maximization in Eq. (\ref{ggmdef}) is over all such  bipartite split ${\cal A}: {\cal B}$.

\noindent\textit{Monogamy of quantum correlations:}
The monogamy relation of a bipartite quantum correlation measure $\mathcal{Q}$, with 
respect to a multipartite quantum state $\rho_{a_1a_2\cdots a_n}$
shared between $n$ parties $a_1$, $a_2$, $\cdots$, $a_n$, demands that \cite{salini,mon_ent,mon_disc} 
\begin{eqnarray}
 \sum_{j\neq k}\mathcal{Q}(\rho_{a_ka_j})\leq\mathcal{Q}(\rho_{a_k:rest}),
 \label{monogamy}
\end{eqnarray}
with $\rho_{a_ka_j}$ being the reduced density matrix of the 
parties $a_k$ and $a_j$ obtained by tracing out all the other parties from
$\rho_{a_1a_2\cdots a_n}$, where $\mathcal{Q}(\rho_{a_k:rest})$ is the value of 
$\mathcal{Q}$ in the bipartition $a_k:rest$, and $j,k\in\{1,2,\cdots,n\}$. This implies that a high value 
of $\mathcal{Q}$ between any two parties of the tripartite 
system does not allow a high value of $\mathcal{Q}$ shared between any one of these parties with any one 
from the rest of the 
parties. The party $a_k$ is referred as ``node'' for the particular quantum correlation measure 
$\mathcal{Q}$. This property is 
observed amongst several of the known quantum correlation measures belonging to both the 
entanglement-separability as well as the 
information-theoretic regimes. The correlation measure that satisfies this property 
is said to be monogamous. For example, for tripartite state, 
squared concurrence as well as squared negativity are monogamous, while quantum discord is not. 

\noindent\textit{Monogamy score:}
One can define a monogamy score from Eq. (\ref{monogamy}) as \cite{score}
\begin{eqnarray}
 \delta_{\mathcal{Q}}(\rho_{a_1a_2\cdots a_n})=\mathcal{Q}(\rho_{a_k:rest})-\sum_{j\neq k}\mathcal{Q}(\rho_{a_ka_j}),
\end{eqnarray}
which is non-negative for all $n$-partite quantum states that satisfy 
Eq. (\ref{monogamy}), and is negative for states that violate 
it. There has been a large number of studies in the direction of 
understanding the monogamy property of quantum correlations 
having both entanglement-separability and information theoretic origin. 
In this paper, we shall focus on the monogamy score for 
squared quantum correlations, namely, concurrence, negativity, quantum discord, and quantum work deficit 
in the case of three-qubit pure states, and we know that except quantum work deficit, all the squared measures 
are monogamous for three-qubit pure states. 

\section*{Three-qubit pure states: GHZ and W class}

\noindent\textit{GHZ class.} The normalized GHZ-class of three-qubit pure states can be represented by 
\cite{ghzstate,zhgstate,dvc}
\begin{equation}
 |\Phi_{ghz}\rangle=\sqrt{K}\left(\cos\alpha|000\rangle+e^{i\beta}\sin\alpha\bigotimes_{i=1}^{3}|\eta_i\rangle\right), 
 \nonumber
\end{equation}
where $|\eta_i\rangle=\cos\gamma_i|0\rangle+\sin\gamma_i|1\rangle$, 
and $K$ being the normalization factor such that  
\begin{eqnarray}
\frac{1}{2}\Big(\frac{1}{K}-1\Big)=\Big(\cos\alpha\sin\alpha\cos\beta\Big)\prod_{i=1}^{3}\cos\gamma_i, 
\end{eqnarray}
and $K \in (1/2,\infty)$. The ranges for the five real parameters are 
$\alpha\in(0,\pi/4]$, $\gamma_i\in(0,\pi/2]$, $i=1,2,3$, and $\beta\in[0,2\pi)$. 

\noindent\textit{W class.} The three-qubit pure states belonging to W class can be parametrized as \cite{W-vs-GHZ,dvc}
\begin{equation}
 |\Phi_{w}\rangle=\sqrt{a}|001\rangle+\sqrt{b}|010\rangle+\sqrt{c}|100\rangle+\sqrt{d}|000\rangle
 \label{state_wclass}
\end{equation}
with $a,b,c>0$, and $d=1-(a+b+c)\geq0$. It is known that a state from the GHZ class can not be transferred to a state from the 
W class by stochastic LOCC (SLOCC) \cite{dvc}.

\end{document}